\documentclass[a4paper]{jpconf}

\usepackage{amsmath,amssymb,hyperref,times,graphicx,bm,bbold,cite}
\def\<{\langle}
\def\>{\rangle}

\newcommand{\Xdag}[2]{\hat{\pmb{#1}}_{#2}^\dag}
\newcommand{\Xd}[2]{\hat{\pmb{#1}}_{#2}}
\newcommand{\cdag}[1]{\Xdag{c}{#1}}
\newcommand{\cd}[1]{\Xd{c}{#1}}
\newcommand{\nop}[1]{\Xd{n}{#1}}

\newcommand{\id}{\mathbb{1}}

\begin{document}

\bibliographystyle{iopart-num}
\title{Entanglement studies of interacting fermionic models}

\author{F Parisen~Toldin and F F Assaad}
\address{Institut f\"ur Theoretische Physik und Astrophysik, Universit\"at W\"urzburg, Am Hubland, D-97074 W\"urzburg, Germany}
\ead{francesco.parisentoldin@physik.uni-wuerzburg.de}
\begin{abstract}
  Recent advances in the field of strongly correlated electron systems allow to access the entanglement properties of interacting fermionic models, by means of Monte Carlo simulations.
  We briefly review the techniques used in this context to determine the entanglement entropies and correlations of the entanglement Hamiltonian.
  We further apply these methods to compute the spin two-point function of entanglement Hamiltonian for a stripe embedded into a correlated topological insulator.
  Further we discuss a recent method that allows an unbiased, numerically exact, direct determination of the entanglement Hamiltonian by means of auxiliary field quantum Monte Carlo simulations.
\end{abstract}

\section{Introduction}
In recent years, a growing number of studies have focused on the quantum entanglement in many-body systems. Its characterization has provided a variety of new insights in old and new problems, complementing the information obtained by more traditional approaches \cite{Laflorencie-16}. In this context, quantum entanglement also plays an important role in Density-Matrix Renormalization Group (DMRG) studies, as it essentially gives a limitation to the various algorithms \cite{Schollwoeck-05}.
The simplest setup to investigate entanglement is the so-called bipartite entanglement, where one divides a system into two parts, conventionally indicated by $A$ and $B$. By tracing out the degrees of freedom of $B$, one obtains the reduced density matrix $\hat{\rho}_A$ of the remaining subsystem $A$. In this framework, the most studied quantities are the von Neumann entropy $S_{\rm vN}$ and especially the Renyi entropies $S_{\rm R}^{(n)}$, defined as
\begin{align}
  S_{\rm vN} &\equiv -\Tr_A \{\hat{\rho}_A \ln \hat{\rho}_A\},
  \label{SvN}\\
  S_{\rm R}^{(n)} &\equiv \frac{1}{1-n} \ln\Tr_A\{\hat{\rho}_A^n\},
  \label{SR}
\end{align}
where $n>1$ is an integer. In the ground state, entanglement entropies generically satisfy an area law, i.e., they are to the leading order proportional to the area between the two subsystems \cite{ECP-10}. Especially of interest are corrections to the area law, which can in principle reveal topological properties \cite{KP-06,LW-06,IHM-11} (see, however, Ref.~\cite{BEB-14}), signal the presence of Goldstone modes \cite{MG-11}, and, most notably, allow to extract the central charge of models described by a 1+1 dimensional Conformal Field Theory (CFT)\cite{CC-04,CC-09}.

Additional information is contained in the entanglement (or modular) Hamiltonian $\hat{H}_E$, defined by\footnote{Strictly speaking, Eq.~(\ref{HE}) defines the modular Hamiltonian, which equals the entanglement Hamiltonian only in the ground state. Here, in agreement with recent literature, we shall make no distinction and always use the term entanglement Hamiltonian.}
\begin{equation}
\hat{\rho}_A \equiv e^{-\hat{H}_E}.
\label{HE}
\end{equation}
The notion of entanglement Hamiltonian is of great interest for several reasons.
The spectrum of $\hat{H}_E$, known as ``entanglement spectrum'', has been shown to display the edge physics of topologically ordered fractional Quantum Hall state \cite{LH-08}, and of symmetry-protected topological states \cite{Fidkowski-10,TZV-10,ALPT-13,Assaad-15}. The entanglement Hamiltonian plays a central role in the so-called first law of entanglement \cite{BCHM-13}.
Moreover, an explicit determination of $\hat{H}_E$ opens the possibility of studying the entanglement using statistical mechanics tools, so as, for instance, to possibly characterize the reduced density matrix as a thermal state. Furthermore, an explicitly known entanglement Hamiltonian could be engineered so as to provide a direct experimental measure of entanglement in many-body systems, an otherwise very difficult task \cite{DVZ-18}.
We also notice that almost all entanglement measures can be obtained from the entanglement Hamiltonian. For example, the expectation value of $\hat{H}_E$ equals the von Neumann entanglement entropy $S_{\rm vN}$, which is generically not accessible in numerical Monte Carlo (MC) simulations; in contrast, $S_{\rm vN}$ can be computed by methods which access to the ground state wave function, such as DMRG or Exact Diagonalization.

Compared to the entanglement entropies, much less results are known for the entanglement Hamiltonian. Its explicit determination has proven to be a considerably more difficult problem,
where only a few solvable results are available.
Besides easily solvable limiting cases, such as in the absence of interactions between the two subsystems, or the high-temperature limit,
a particularly important case is that of a relativistic field theory with Hamiltonian density $\hat{H}(x)$ in flat $d-$dimensional Minkowski space. The entanglement Hamiltonian for a semi-infinite subspace $A=\{(x_1,\ldots,x_d) \in \mathbb{R}^d, x_1>0\}$ is\cite{BW-75,BW-76}
\begin{equation}
\hat{H}_E = 2\pi \int_A d^dx \left(x_1 \hat{H}(x)\right) + c,
\label{BW}
\end{equation}
where $c$ is a constant. The right-hand side of Eq.~(\ref{BW}) is known as Bisognano-Wichmann (BW) form of the entanglement Hamiltonian, or Rindler Hamiltonian \cite{Susskind_book}.
 With additional conformal symmetry, a mapping of $A$ to a ball allows again to express $\hat{H}_E$ as an integral of $\hat{H}(x)$ \cite{CHM-11}.
Ref.~\cite{CT-16} reviews the cases in $1+1$ dimensional CFT, where $\hat{H}_E$ can be expressed as an integral.
On the lattice, a few cases are exactly known. For a noncritical one-dimensional transverse-field Ising model, and for the XXZ model in the massive phase, the entanglement Hamiltonian for a semi-infinite line subsystem has been exactly determined \cite{IT-87,PKL-99}. For a free (nonrelativistic) fermionic system the entanglement Hamiltonian is\cite{Peschel-04}
\begin{equation}
  \hat{H}_E = \ln\left[\left(G_A^T\right)^{-1}-\id\right] - \ln\det(\id-G_A),\qquad (G_A)_{ij}\equiv\<\cdag{i}\cd{j}\>,
  \label{HEfree}
\end{equation}
where $\cdag{i}$, $\cd{j}$ are the fermionic creation and annihilation operators in the subsystem $A$, and $G_A$ is the Green's function matrix of the model, restricted to $A$. Despite the exact result of Eq.~(\ref{HEfree}), an explicit computation of the entanglement Hamiltonian in the deceptively simple case of a free fermionic chain, and for a subsystem consisting in a segment, has eluded an analytical treatment so far. In this case, even a numerical computation of Eq.~(\ref{HEfree}) is technically hard, because $G_A$ is plagued by eigenvalues close to $0$ and $1$, rendering the formula of Eq.~(\ref{HEfree}) numerically unstable.
The lattice models mentioned so far share  a common property of being described by a CFT in the low-energy limit. Thus, one may conjecture that the corresponding entanglement Hamiltonian attains the BW form, up to a lattice discretization. Indeed, this is the case for the aforementioned exact results. Nevertheless, for a free fermionic chain, the entanglement Hamiltonian for a segment displays intriguing corrections to the BW form which, remarkably, persist even in the limit of a long segment \cite{EP-17}. Recent studies have provided further numerical evidence in support of a lattice-discretized BW form of the entanglement Hamiltonian for various one- and two-dimensional lattice models \cite{KKTH-16,DVZ-18,KLC-18,ZHH-18,GMSCD-18}.
In Ref.~\cite{PTA-18} we have introduced a method which allows a numerically exact determination of the entanglement Hamiltonian for generic interacting fermionic models, exploiting the auxiliary field Quantum Monte Carlo (QMC) method \cite{Blankenbecler81,White89,AF_notes}.

In this paper we focus on the QMC studies of entanglement in interacting fermionic models. In Sec.~\ref{sec:entropies} we review the techniques used to compute the Renyi entropies. In Sec.~\ref{sec:HEcorr} we present some results for the correlations of the entanglement Hamiltonian in a correlated topological insulator. In Sec.~\ref{sec:HE} we briefly discuss the method introduced in Ref.~\cite{PTA-18} to explicitly determine the entanglement Hamiltonian. We conclude in Sec.~\ref{sec:outlook} with an outlook to future research topics.

\section{Entanglement entropies}
\label{sec:entropies}
\begin{figure}[t]
  \includegraphics[width=0.4\linewidth]{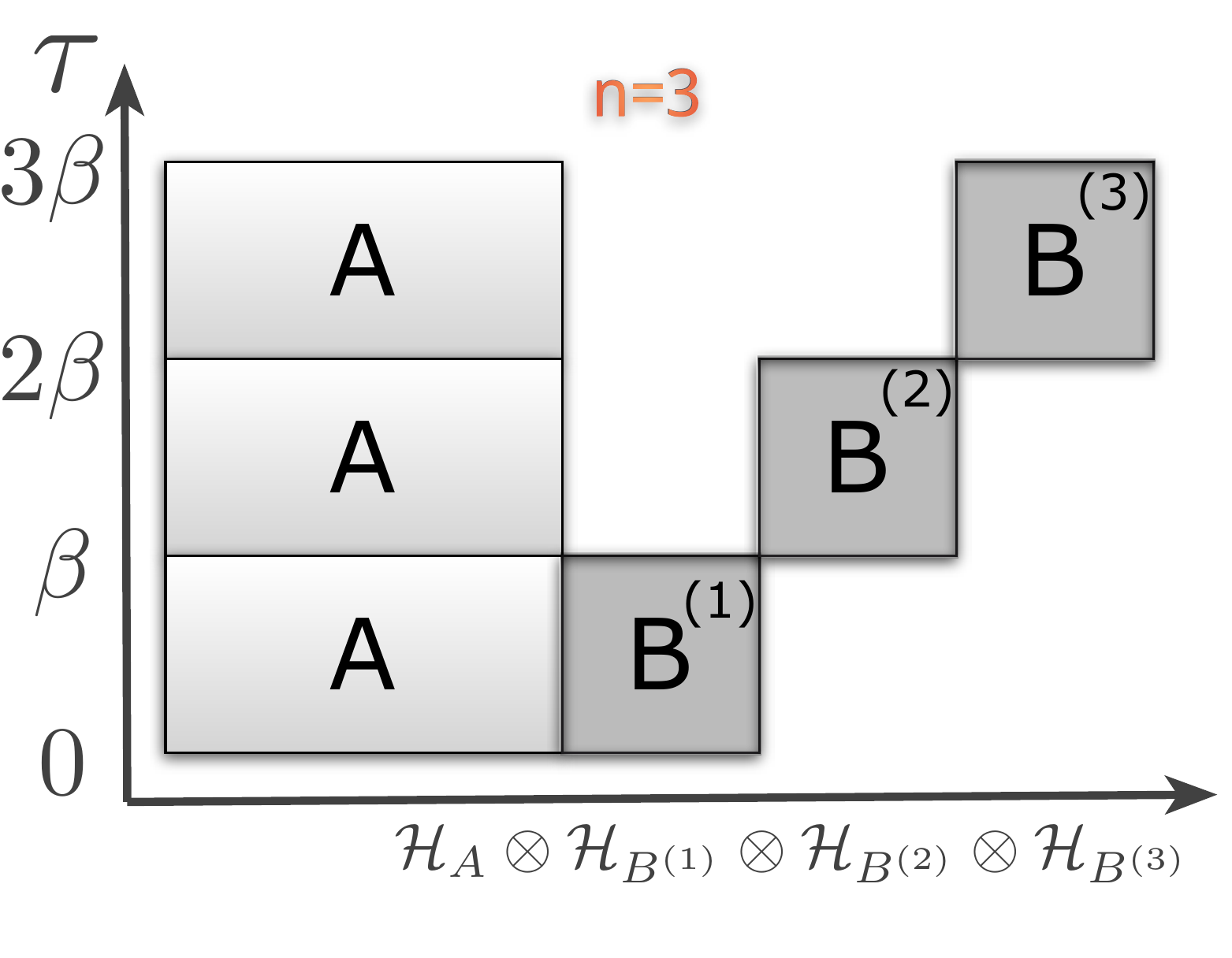}
  \hspace{0.05\linewidth}
  \begin{minipage}[b]{0.5\linewidth}
    \caption{\label{replica}Illustration of the replica trick for $n=3$.
      The effective time-dependent Hamiltonian vanishes in the white regions.}
  \end{minipage}
\end{figure}
The auxiliary field QMC method is a widely used MC method for fermionic models \cite{Blankenbecler81,White89,AF_notes}. It can be conveniently programmed using the ALF package \cite{ALF}.
Its formulation relies in a Trotter decomposition, followed by a Hubbard-Stratonovich (HS) transformation, which decouples the interactions via HS fields $\{s\}$. The MC consists then in a stochastic sampling of the resulting probability distribution $P(\{s\})$ of the HS fields. At each $\{s\}$ configuration the system corresponds to a free fermionic model, hence by using Eq.~(\ref{HEfree}) one obtains an expression for the reduced density matrix $\hat{\rho}_A$\cite{Grover-13}
\begin{equation}
  \hat{\rho}_A = \int d\{s\} P(\{s\}) \det\left(\id - G_A(\{s\})\right) e^{-\cdag{i}h_{ij}(\{s\})\cd{j}}, \qquad
  h(\{s\}) = \ln\left( (G_A(\{s\})^T)^{-1} -\id \right),
  \label{rhoAF}
\end{equation}
where now the restricted Green's function matrix $G_A(\{s\})$ explicitly depends on the HS fields configuration. To compute the $n^{\rm th}$ Renyi entropy, one introduces $n$ parallel simulations with HS fields $\{s_i\}$. Using Eq.~(\ref{SR}), $S_{\rm R}^{(n)}$ is obtained by sampling an observable depending on all $\{s_i\}$:
\begin{equation}
  e^{-(n-1)S_{\rm R}^{(n)}} = \int \prod_{i=1}^n d\{s_i\} P(\{s_i\}) \left[\prod_{i=1}^n\det\left(\id-G_A(\{s_i\})\right)\right] \det\left(1+\prod_{i=1}^n\frac{G_A(\{s_i\})}{\id-G_A(\{s_i\})}\right).
  \label{SRAF}
\end{equation}
Using Eq.~(\ref{SRAF}), Renyi entropies have been computed in Refs.~\cite{Grover-13,ALPT-13,DP-15,DP-16}. While this approach is simple compared to other methods, it suffers from two main issues. The first one is that the sampled observable appears to be affected by considerably large fluctuations, increasing with the interaction strength, such that $S_{\rm R}^{(n)}$ is easily computed in the weak-coupling regime only \cite{ALPT-13,DP-15}.
A solution to this issue, obtained within the framework of the Hybrid QMC, where also Eq.~(\ref{rhoAF}) holds, consists in reformulating Eq.~(\ref{SRAF}) as an integral over an auxiliary parameter of an observable sampled with a modified measure \cite{DP-15,DP-16}.
The replica method discussed below provides another solution to the problem.
The second issue in using Eq.~(\ref{SRAF}) is due to the fact that, in sampling the ground state, $G_A(\{s_i\})$ shows eigenvalues close to $1$, giving a singularity on the right-hand side of Eq.~(\ref{SRAF}) for $n>2$.
(For $n=2$ Eq.~(\ref{SRAF}) can be easily simplified, avoiding the singularity \cite{Grover-13}.)
The singularity can be avoided by introducing a thermal broadening, as we discuss in Ref.~\cite{ALPT-13}, or by recasting the right-hand side of Eq.~(\ref{SRAF}) as a determinant of a bigger matrix \cite{DP-16}.

An alternative, more general, framework to study the entanglement is the so-called replica trick.
It has been used in fermionic \cite{BT-14,WT-14,Assaad-15,BT-16}, bosonic \cite{IHM-11}, spin systems \cite{HGKM-10,HR-12}, and it is also a starting point for a field-theory approach \cite{CC-04,CC-09}. This method consists in a representation of $\hat{\rho}_A^n=\exp(-n\hat{H}_E)$, obtained by introducing $n$ replicas of the subsystem $B$,
enlarging the original Hilbert space ${\cal H}={\cal H}_A\otimes{\cal H}_B$ to ${\cal H}_R = {\cal H}_A\otimes{\cal H}_{B^{(1)}}\otimes\ldots\otimes{\cal H}_{B^{(n)}}$.
In ${\cal H}_R$ one defines a time-dependent Hamiltonian, such that the subsystem $A$ evolves in the imaginary-time interval $[0,n\beta]$, whereas each replica evolves in an interval $[i\beta,(i+1)\beta]$.
An illustration of the replica trick for $n=3$ is shown in Fig.~\ref{replica}.
The replica trick, combined with the swap algorithm \cite{HR-12}, allows to compute the Renyi entropies in fermionic models, avoiding the aforementioned sampling issues \cite{BT-14}, but with an increased computational cost $\propto n^3$.
As we discuss in Ref.~\cite{Assaad-15}, the approach based on Eq.~(\ref{SRAF}) and the replica trick are formally equivalent.

\section{Correlations of the Entanglement Hamiltonian}
\label{sec:HEcorr}
By sampling a Gibbs measure $\sim \exp(-n\hat{H}_E)$, the replica trick allows to compute the correlations of the entanglement Hamiltonian at an effective integer inverse temperature $n$. It is important to note that the entanglement Hamiltonian also depends implicitly on the physical inverse temperature $\beta$. (See Eq.~(\ref{HE})).
We apply this technique to the Kane-Mele-Hubbard model, a paradigmatic model to study the interplay of topology and correlations \cite{ABH-13,BHA-14,HA-13,HLA-11,HMLWMA-12,HPTHA-14,KM-05b,KM-05,PTHAH-14}. The model is defined on a honeycomb lattice, with Hamiltonian
\begin{equation}
  \hat{H} = -t\sum_{\<i\ j\>,\sigma}\cdag{i,\sigma}\cd{j,\sigma} +i\lambda\sum_{\<\<i\ j\>\>}\cdag{i}(\vec{\nu}_{ij}\cdot\vec{\sigma})\cd{j} + U\sum_i\left(\nop{i,\uparrow}-\frac{1}{2}\right)\left(\nop{i,\downarrow}-\frac{1}{2}\right),
\label{KM}
\end{equation}
with $\cdag{i}\equiv(\cdag{i,\uparrow}, \cdag{i,\downarrow})$, $\cd{i}\equiv(\cd{i,\uparrow}, \cd{i,\downarrow})$ the creation and annihilation operator vectors for nonrelativistic spinful fermions, $\nop{i,\sigma}\equiv\cdag{i,\sigma}\cd{i,\sigma}$ the number operator, and $\vec{\sigma}$ the vector of Pauli matrices. In Eq.~(\ref{KM}) the first term represents hoppings between nearest-neighbor sites, the second term couples next-nearest-neighbors sites and represents the spin-orbit interaction, and the third term is a on-site Hubbard repulsion. The unit vectors $\vec{\nu}_{ij}$ take values $\vec{\nu}_{ij}=\pm\vec{e}_z$ depending on the sublattice, spin, and direction of hoppings; see, e.g., Ref.~\cite{PTHAH-14} for a definition. At half-filling the Kane-Mele-Hubbard Hamiltonian can be simulated using the auxiliary field QMC without encountering the sign problem. For $\lambda=0$ the model reduces to the Hubbard model and exhibits a quantum phase transition in the ground state at $U/t=3.80(1)$\cite{PTHAH-14} in the Gross-Neveu-Heisenberg universality class\cite{Herbut-06,HJR-09,HJV-09,PTHAH-14}. For $\lambda>0$ the model shows a quantum phase transition in the classical 3D XY universality class, between a quantum spin-Hall insulator phase at small values of $U/t$ and an antiferromagnetic insulating phase at large values of $U/t$ \cite{HLA-11,HLA-11_erratum,HMLWMA-12,PTHAH-14}.
\begin{figure}[t]
  \begin{minipage}[t]{0.45\linewidth}
    \includegraphics[width=\linewidth]{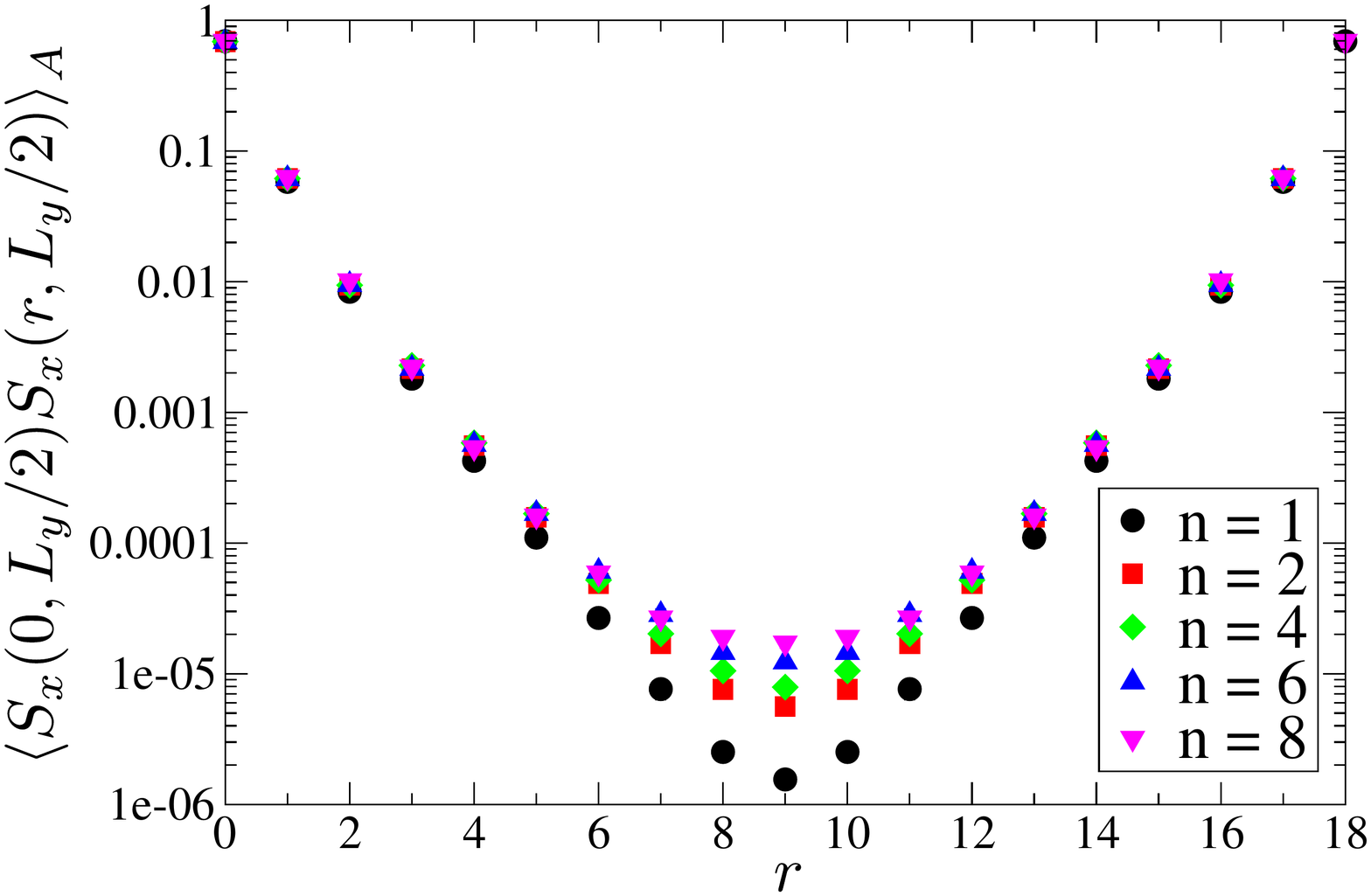}
    \caption{\label{corr_middle}Semilogarithmic plot of the spin-spin correlations of the entanglement Hamiltonian in the middle of a stripe.}
  \end{minipage}
  \hspace{0.05\linewidth}
  \begin{minipage}[t]{0.45\linewidth}
    \includegraphics[width=\linewidth]{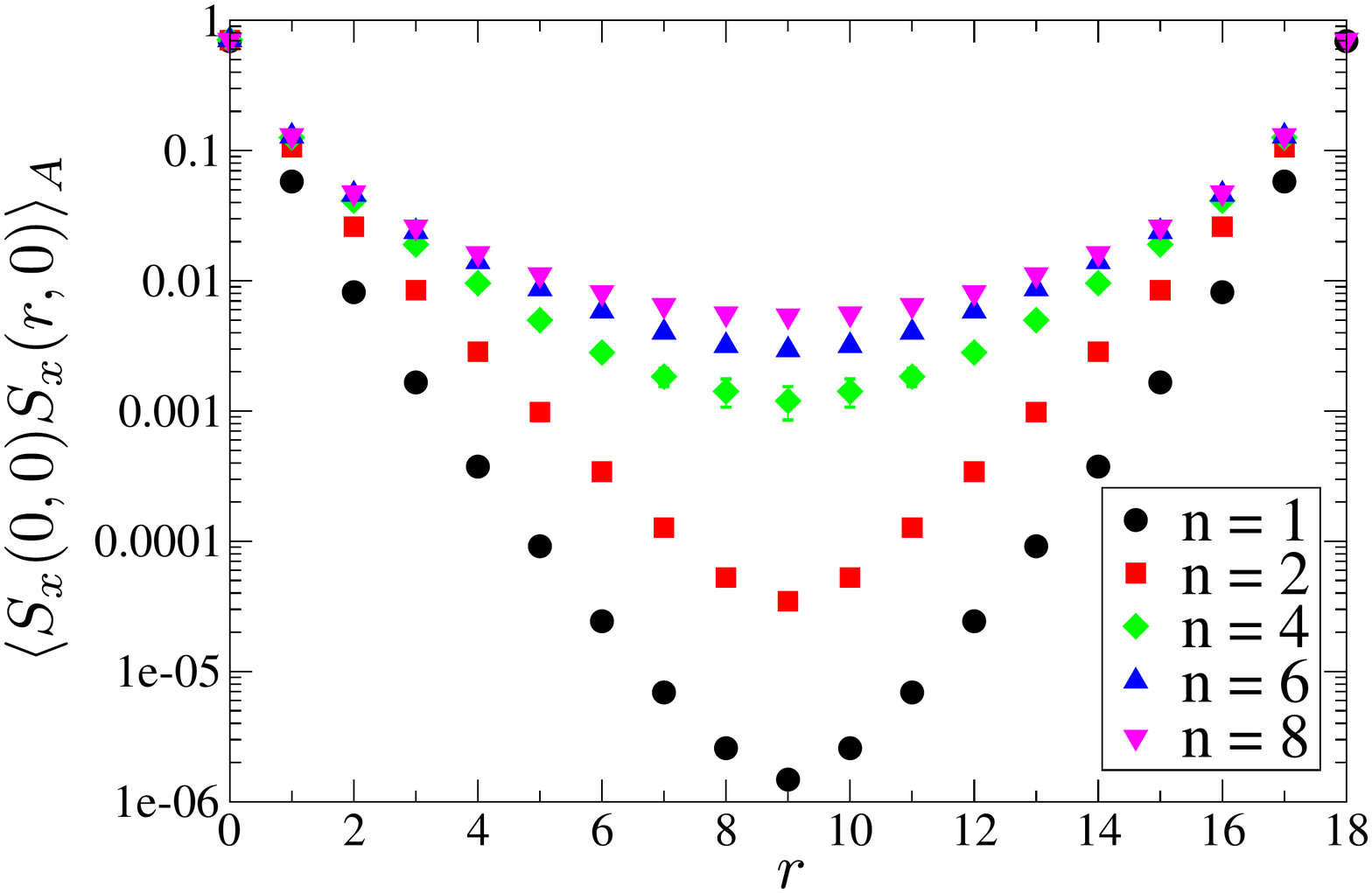}
    \caption{\label{corr_edge}Same as Fig.~\ref{corr_middle} for spin-spin correlations along the edge of the stripe.}
  \end{minipage}
\end{figure}

We consider the model in the ground state for $U/t = 4$ and $\lambda/t=0.2$, in the quantum spin-Hall phase, and compute the spin-spin correlations of the entanglement Hamiltonian for a stripe of width $W_A=3$, translationally invariant in one direction, cut inside a finite $9\times 18$ lattice with periodic boundary conditions. In Fig.~\ref{corr_middle} and Fig.~\ref{corr_edge} we show the results for the spin-spin correlations in the middle of the stripe, and along its edge, for various values of the replica index $n$. For $n=1$ by definition the correlations of $\hat{H}_E$ are identical to those of $\hat{H}$, and show the expected exponential decay of the gapped spin-Hall insulating phase. Upon increasing $n$ the correlations in the middle do not change qualitatively, whereas those on the edge of the strip display a slower decay. For $n=8$ we approach the ground state properties of the entanglement Hamiltonian: while the correlations in the middle of the strip are still exponentially decaying, those on the edge support a power-law decay, analogous to the gapless modes
found at the physical edge of a correlated quantum spin-Hall insulator \cite{HA-13}. A similar correspondence is found
by studying the imaginary-time displaced Green's function of
the entanglement Hamiltonian \cite{Assaad-15}.
Time-displaced correlation functions of the entanglement Hamiltonian can
be also used to compute the low-lying entanglement spectrum  \cite{ALPT-13,Assaad-15}.

\section{Determination of the Entanglement Hamiltonian}
\label{sec:HE}
Eq.~(\ref{rhoAF}) is the starting point for a method that we have introduced in Ref.~\cite{PTA-18} to explicitly determine the entanglement Hamiltonian for interacting fermionic models, within the framework of auxiliary field QMC simulations. We briefly describe here the main ideas of the method. The exponential appearing on the right-hand side of Eq.~(\ref{rhoAF}) admits an exact expansion in terms of normal-ordered many-body operators, whose coefficients can be sampled using the auxiliary field QMC, obtaining a MC determination of the reduced density matrix $\hat{\rho}_A$. We then compute the negative logarithm of the sampled $\hat{\rho}_A$, and express it in terms of fermionic creation and annihilation operators. This results in an unbiased, numerically exact determination of the entanglement Hamiltonian, as a sum of normal-ordered operators. We refer to Ref.~\cite{PTA-18} for more details on the method.

We have applied the technique to the Hubbard chain model, obtaining the one-body term of $\hat{H}_E$ for a segment of length $L=8$ embedded in a chain of size $L=32$, and at inverse temperatures $1\le\beta/t\le 3$. The resulting nearest-neighbor hopping terms in $\hat{H}_E$ qualitatively resemble the BW form, while other hopping terms
are negligible.
In Ref.~\cite{PTA-18} we have also determined the entanglement Hamiltonian for a two-legs Hubbard chain.
This model exhibits a single gapped phase \cite{NWS-96,WOHB-01}, ensuring a fast approach to the thermodynamic limit in the linear size \cite{Neuberger-89,WAPT-17}. The entanglement Hamiltonian
for a single leg
shows an interesting temperature dependence. At high temperatures $\hat{H}_E$ matches a Hubbard Hamiltonian, in line with general arguments. Upon lowering the temperature $\hat{H}_E$ exhibits a crossover to a Heisenberg-like Hamiltonian, characterized by a large $U/t$ ratio, and the emergence of additional nearest-neighbors antiferromagnetic and next-nearest-neighbors ferromagnetic spin-spin interactions. This result resembles what is found in the two-legs antiferromagnetic Heisenberg model upon tracing out one leg \cite{Poilblanc-10,CPSV-11,PC-11,LS-12,SL-12}.

\section{Outlook}
\label{sec:outlook}
Recent progresses in the entanglement studies of correlated fermionic systems
form a basis for promising future research. The established techniques for entropies can be used to compute other entanglement measures, such as the mutual information. At the same time, investigations of the entanglement Hamiltonian via its correlations, together with its explicit determination, can foster a fruitful exchange with statistical mechanics and field theory approaches, enabling a deeper understanding of correlated many-body systems.

\ack
FPT  thanks the German Research Foundation (DFG)  through the grant No. AS120/13-1  of the FOR 1807.   FFA  thanks  the DFG for financial support through the SFB 1170 ToCoTronics. We acknowledge the computing time granted by the John von Neumann Institute for Computing (NIC) and provided on the supercomputer JURECA \cite{Jureca16} at the J\"ulich Supercomputing Center.
\section*{References}
\bibliography{francesco,fassaad}

\end{document}